\documentclass[preprint, showpacs, preprintnumbers,amsmath,amssymb,nofootinbib]{revtex4}

\usepackage{epsf}
\usepackage{graphicx}

\newcommand{\be}[1]{\begin{equation}\label{#1}}
 \newcommand{\ee}{\end{equation}}
 \newcommand{\bea}{\begin{eqnarray}}
 \newcommand{\eea}{\end{eqnarray}}

 \def\gsim{ \lower .75ex \hbox{$\sim$} \llap{\raise .27ex \hbox{$>$}} }
 \def\lsim{ \lower .75ex \hbox{$\sim$} \llap{\raise .27ex \hbox{$<$}} }

\begin{document}
 \title{
 Inverse volume
corrections to emergent tachyonic inflation in loop quantum
cosmology}

\author{ Puxun Wu\;$^{1,2}$\footnote{wpx0227@gmail.com}, Shuang Nan Zhang\;$^{1,3,4}$ and Hongwei Yu\;$^{2}$}

\address
{$^1$Department of Physics and Tsinghua Center for Astrophysics,
Tsinghua University, Beijing 100084, China
\\
$^2$Department of Physics and Institute of  Physics, Hunan Normal
University, Changsha, Hunan 410081, China
\\
 $^3$Key Laboratory of Particle Astrophysics, Institute of
High Energy Physics, Chinese Academy of Sciences, P.O. Box 918-3,
Beijing 100049, China
\\
$^4$Physics Department, University of Alabama in Huntsville,
Huntsville, AL 35899, USA }

\begin{abstract}
The emergent model in the context of loop quantum
cosmology with a tachyon scalar field is studied. We find that there
is a center equilibrium point in the semiclassical region and a
saddle point in the classical region. If the potential of the
tachyon field satisfies some conditions, the universe can stay at
the center equilibrium point past-eternally and then oscillate
infinitely around this point with the tachyon climbing up its
potential. Once the potential reaches a critical value, these two
equilibrium points coincide with each other and the oscillation
phase is broken by an emergent inflation. In order to obtain a
successful emergent tachyon inflation, a constraint on
$\dot{\phi}^2$ of tachyon is required.
\end{abstract}


 \maketitle

\section{Introduction}
The inflationary model is very successful to solving some problems
in the standard cosmological model and is consistent with
observations of the Cosmic Microwave Background radiation and high
redshift surveys~\cite{Peiris2003}.  However, the existence of a big
bang singularity in the early universe is still an unresolved
problem. Some authors have tried to address this problem within the
framework of quantum gravity and have suggested some models to avoid
this singularity, such as the pre-big bang~\cite{Gasperini2003,
Lidsey2000} and ekpyrotic/cyclic scenarios~\cite{Khoury2001} in
string/M-theory, but it still remains unclear as to what process
could lead to a nonsingular transition from the pre- to post big
bang phase.

Recently, in order to establish a singularity-free inflationary
model in the context of classical general relativity, a new
scenario, an emergent universe~\cite{Ellis2004a, Ellis2004b}, has
been proposed. In this model, the space curvature is positive and
the universe stays past-eternally in an Einstein static state  and
then evolves into a subsequent inflationary phase. Thus the big bang
singularity can be avoided. In addition, researches show that the
entropy considerations favor the Einstein static state as the
initial state of our universe ~\cite{Gibbons1988}. However the
Einstein static universe in the classical general relativity is
unstable. Therefore the  universe is extremely difficult to maintain
such an initial static state in a long time due to the existence of
perturbations, such as the quantum fluctuations.

More recently, the emergent model within the framework of quantum
gravity has been studied extensively. In this regard, Mulryne et
al.~\cite{Mulryne2005} studied the existence and stability of
Einstein static state in the context of Loop Quantum
Cosmology~(LQC)~\cite{LQC, LQC2} with the inflaton scalar field
modified by the inverse volume corrections in Loop Quantum
Gravity~(LQG) (see \cite{LQG} for recent reviews), where LQC is an
application to cosmology of LQG. The inverse volume is the cube of
inverse scale factor and in LQC the correction to the inverse volume
arises  by exploiting the ambiguity in defining the inverse volume
operator which is required to quantize the Hamiltonian
constraint~\cite{Bojowald2001}. It was found that there is a center
equilibrium point in the semiclassical region besides a saddle point
in the classical region. If  the potential energy of scalar field is
less than a critical value, the universe can stay past-eternally at
a static state and then enter into an oscillating phase with  the
scalar field being driven up its potential~\cite{Lidsey2004}. As a
result, once the potential reaches this critical value, the center
equilibrium point coincides with the saddle one, and the oscillating
phase is broken and then the universe enters an inflation era. Thus
a successful singularity-free inflation model is obtained.  However
in Ref.~\cite{Mulryne2005}, the author did not consider the inverse
volume correction to the gravity, which seems to be ad-hoc. The
reason is that  the same ambiguity also arises in the gravitational
part of the Hamiltonian constraint. Therefore the inverse volume
correction to the gravity should be considered, which may change the
dynamics significantly.  For the case of the LQC without the inverse
volume modification, although there is also a new stable center if
the cosmological constant is larger than a critical scale, the
universe cannot  break the infinite cycles around this center point
to naturally enter a subsequent inflationary
phase~\cite{Parisi2007}. Thus it will be interesting to study the
emergent inflation in LQC with the inverse volume corrections both
to matter and gravity. In addition  Lidsey and
Mulryne~\cite{Lidsey2006} studied the dynamics of a scalar field
within the context of branewold scenario proposed by Shtanov and
Sahni~\cite{Shtanov2003, Shtanov2002} and found that, with the
scalar field climbing up its potential, the cosmic evolution from
the past eternal cycle  to inflation can occur naturally as in the
case of LQC with the scalar field modification. Later, modified
stabilities of the Einstein static model in some non-Einstein 
gravities~\cite{Boehmer2007} have also been discussed.

In this paper, we will discuss an emergent inflation in LQC with the
inverse volume corrections to both the matter and gravity. A tachyon
scalar field is considered since it has a non-canonical kinetic term
and has attracted considerable attention for the cosmic inflation at
early times and for dark energy and dark matter at late
times~\cite{Sen2002}. In the framework of LQC the tachyon inflation
has been studied in the cases of the matter with or without the
inverse volume correction and results showed that the super
inflation can appear easily in both cases~\cite{Sen2006, Xiong2007}.

\section{Tachyon in LQC}
In a positively curved  Friedmann-Robertson-Walker (FRW) background,
following the calculations in Refs.~\cite{Hami, Vandersloot}, the
effective Hamiltonian for the LQC system with tachyon field can be
expressed \bea \mathcal{H}_{eff}=-\frac{3}{\kappa \gamma^2} S_J(p)
\frac{\sin^2(\overline{\mu} c)}{\bar{\mu}^2} -\frac{3S_J(p)
V_0^{2/3}}{\kappa}+p^{3/2}\rho_\phi\;,\eea with \bea S_J(p)=
\sqrt{p}S(q)\;,\eea where $\kappa=8\pi/M_{pl}^2$, $p=a^2$, $J$ is a
half-integer, $\rho_\phi$ is the energy density of tachyon field,
$q=(\frac{a}{a_*})^2$ with $a_*=(8\pi \alpha^{1/2} \gamma J)^{1/3}
l_{pl}$ and $\gamma=0.2375$. In Eq. (2), $S(q)$ arises from the
inverse volume correction to gravity  and can be expressed as\bea
S(q)=\frac{4}{\sqrt{q}}\bigg\{\frac{1}{10}\big[(q+1)^{5/2}+\textrm{sign}(q-1)|q-1|^{5/2}\big]
-\frac{1}{35}\big[(q+1)^{7/2}-|q-1|^{7/2}\big]\bigg\}. \eea   Using
the Hamilton's equation we can obtain the modified Friedmann
equation \bea\label{H2}
H^2=\big(\frac{\kappa}{3}\rho_\phi-\frac{S}{a^2}\big)\big(S+\frac{3S}{\kappa\rho_c
a^2}-\frac{\rho_\phi}{\rho_c}\big)\;,\eea where $\rho_c\simeq
0.82M_{pl}^4$ is the critical LQC energy density. When $S=1$, the
above Friedmann equation  corresponds to the case without the
inverse volume correction in the positively curved universe given in
Ref.~\cite{Parisi2007}. Since $H^2$ must be non-negative, we have
the following limits \bea
\frac{3}{\kappa}\frac{S}{a^2}\leq\rho_\phi\leq
\rho_c\bigg(S+\frac{3S}{\kappa\rho_c a^2}\bigg)\;.\eea

In LQC with the inverse volume correction  the energy density and
pressure for tachyon matter can be written as~\cite{Sen2006}
\bea\label{rho}
\rho_\phi=\frac{Vp^{3/2}|F(q)|^{3/2}}{\sqrt{p^{3}|F(q)|^{3}-\dot{\phi}^2}}\;,\eea
\bea\label{p}
p_\phi=-\frac{Vp^{3/2}|F(q)|^{3/2}}{\sqrt{p^{3}|F(q)|^{3}-\dot{\phi}^2}}
\bigg[1+\frac{\dot{\phi}^2|F(q)|'}{p^2|F(q)|^4}\bigg]\;.\eea Here
$V$ is the potential of tachyon field, which must be larger than
zero since the energy density must be positive,  and \bea
F(q)&=&a_*^{-2}\bigg(\frac{8}{77}\big[7\{(q+1)^{11/4}-|q-1|^{11/4}\}\nonumber\\
&&-11q\{(q+1)^{7/4}-\textrm{sign}(q-1)|q-1|^{7/4}\}\big]\bigg)^{4}\;,\eea
By differentiating Eq.~(\ref{H2}) with time  and using
Eqs.~(\ref{rho}, \ref{p}) we have \bea\label{DH}
\dot{H}&=&\frac{1}{2}\bigg(3\kappa \rho_\phi
\bigg(1-\frac{V^2}{\rho_\phi^2}\bigg)\frac{d \ln F}{d \ln
q}+{2S\over a^2}-{2q\over a^2}{dS\over d
q}\bigg)\bigg(S+\frac{3S}{\kappa\rho_c
a^2}-\frac{\rho_\phi}{\rho_c}\bigg)\\
\nonumber
&&+\bigg(\frac{\kappa}{3}\rho_\phi-\frac{S}{a^2}\bigg)\bigg(2q\frac{dS}{dq}+\frac{6q}{\kappa\rho_c
a^2}\frac{d S}{d q}- \frac{6 S}{\kappa\rho_c
a^2}-\frac{3\rho_\phi}{\rho_c}\bigg(1-\frac{V^2}{\rho_\phi^2}\bigg)\frac{d
\ln F}{d \ln q}\bigg)\eea In the following a constant potential is
considered which is a good approximation if the variation of
potential is very slow with time.

\begin{figure}[htbp]
\includegraphics[width=7cm]{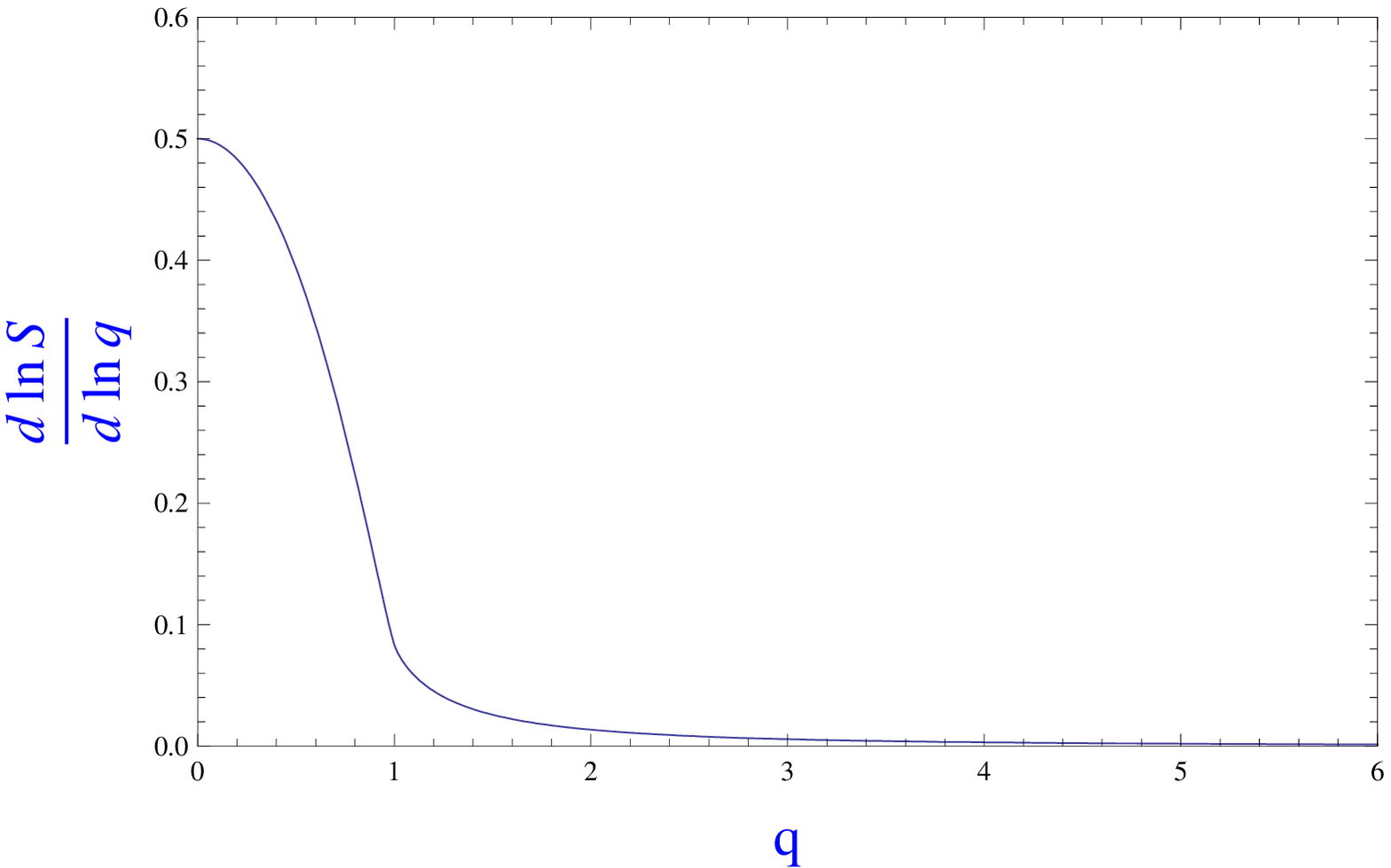}
 \caption{\label{Fig1} The evolutionary curves of function $\frac{d\ln S}{d\ln q}$ against $q$ . }
\end{figure}

The equilibrium points of this system are given by the conditions
$\ddot{a}=\dot{a}=0$, which imply \bea a=a_{eq}\;, \qquad
H(a_{eq})=0.\eea From Eq.~(\ref{H2}) it is easy to obtain  that
$H^2=0$ corresponds to two critical energy densities. We find that
the critical density $\rho= \rho_c\big(S+\frac{3S}{\kappa\rho_c
a^2}\big)$ cannot lead to a successful emergent inflation. Therefore
in the following we will only consider the critical point
$\rho_\phi=\frac{3S}{\kappa a^2}$. From Eq.~(\ref{DH}), one can
obtain the following constraint equations for $a_{eq}$
 \bea A(a_{eq})=B(a_{eq}),\qquad B(a)\equiv \frac{\frac{2q}{a^2}\frac{dS}{dq} - \frac{2
 S}{ a^2}}{3\kappa\rho_\phi\big(1-\frac{V^2}{\rho_\phi^2}\big)}\;.\eea
Here $A(a)=\frac{d \ln F}{d \ln q}$. Clearly the condition for the
existence of the equilibrium points is that the functions $A(a)$ and
$B(a)$ intersect. Using Eq.~(\ref{rho}) at the equilibrium point we
find
 \bea \dot{\phi}^2=\frac{2}{9}p^3 |F|^3\frac{\frac{d\ln S}{d\ln q}-1}{B}.\eea
 Since $\frac{d\ln
S}{d\ln q}$ is smaller than $1$ as shown in Fig.~(\ref{Fig1}), the
reality condition, $\dot{\phi}^2>0$, requires that the intersections
of functions $A$ and $B$ must appear in the range of $B<0$.

\begin{figure}[htbp]
\includegraphics[width=7cm]{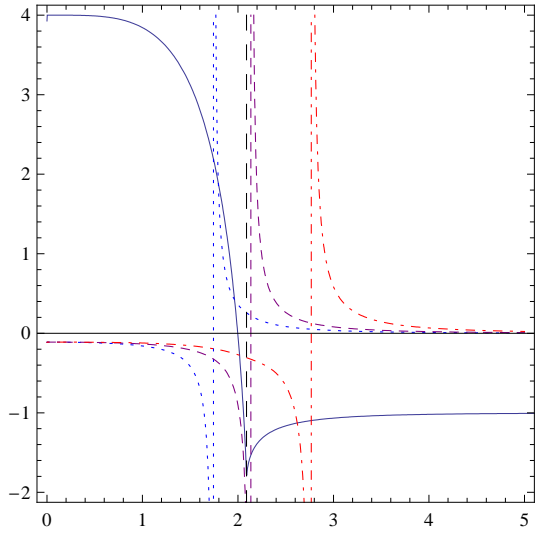}
 \caption{\label{Fig2} The evolutionary curves of functions $A$ (solid line) and
$B$ (dotted, dashed and dot-dashed lines) against $a$ with Plank
unit. The dotted, dashed, and   dot-dashed curves correspond to the
function $B$ with $V>V_{crit}$, $V=V_{crit}$ and $V<V_{crit}$,
respectively. The vertical long-dashed line denotes the position of
$a_*$. }
\end{figure}

In  Fig.~(\ref{Fig2})  we give the curves of functions $A$ and $B$.
The function $A$ reaches its maximum $A_{max}=4$ at $a=0$, decreases
to its minimum $A_{min}=-\frac{11}{6}$ at $a=a_*$, and then
asymptotes to $-1$ at $a\rightarrow \infty$. The function $B$ is a
hyperbola with a single vertical asymptote given by solving
$V=\frac{3S}{\kappa a^2}$. However, from the requirement of
$\dot{\phi}^2>0 $,  only the down-left branch of this hyperbola
plays a role in determining the existence of the equilibrium points.

\begin{figure}[htbp]
\includegraphics[width=7cm]{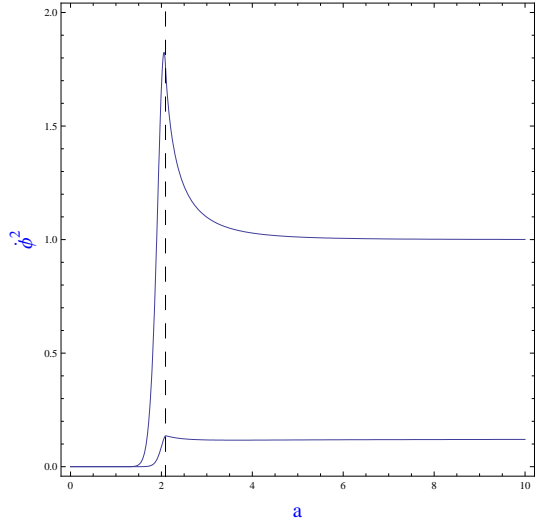}
\caption{\label{Fig3} The constraint on $\dot{\phi}^2$ given in
Eq.~(\ref{phi2}) with Plank unit.  The vertical long-dashed line
denotes the position of $a_*$.}
\end{figure}

If the potential satisfies the conditions $0<V<V_{crit}$ where
$V_{crit}=\frac{3S}{\kappa
a^2}\sqrt{\frac{29}{33}+\frac{4a^2}{33}\frac{d\ln S}{d \ln
q}}|_{a=a_*}$ obtained from the equation $A_{min}=B$, there are two
intersects between the curve of $A$ and the down left branch of $B$.
As shown in Fig.~(\ref{Fig2}), one equilibrium point occurs at the
semiclassical region; the other at the classical region. With the
increasing of the potential these two intersects become closer and
closer. Once $V=V_{crit}$ they coincide with each other and there is
only an equilibrium point. In addition, we find that, in order to
obtain two intersect points of functions $A$ and $B$,  the kinetic
term $\dot{\phi}^2$ of tachyon field must satisfy the following
constraint \bea\label{phi2} \frac{4}{33}|F|^3a^6\big(1-a^2\frac{d\ln
S}{d\ln q}\big)<\dot{\phi}^2<a^6|F|^3\;,  \eea   which comes from
the requirement $0<V< V_{crit}$.  In Fig.~(\ref{Fig3}) we give the
allowed region of $\dot{\phi}^2$. From this Figure, we can see that
$a_{eq}$ cannot be much smaller than $a_*$ and the region of
$a_{eq}\lesssim 0.7 a_*$ is forbidden.

\begin{figure}[htbp]
\includegraphics[width=7cm]{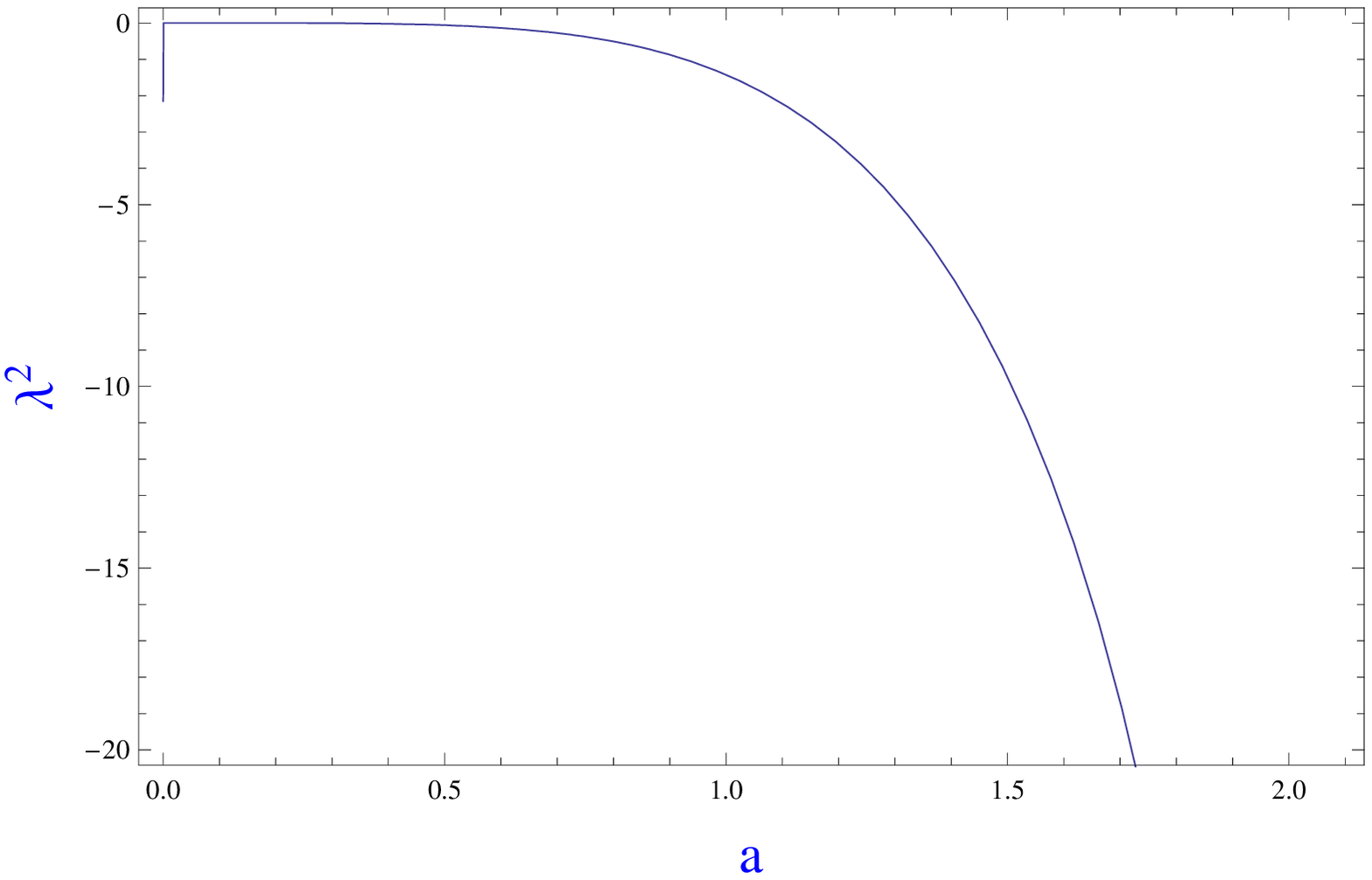}\includegraphics[width=7cm]{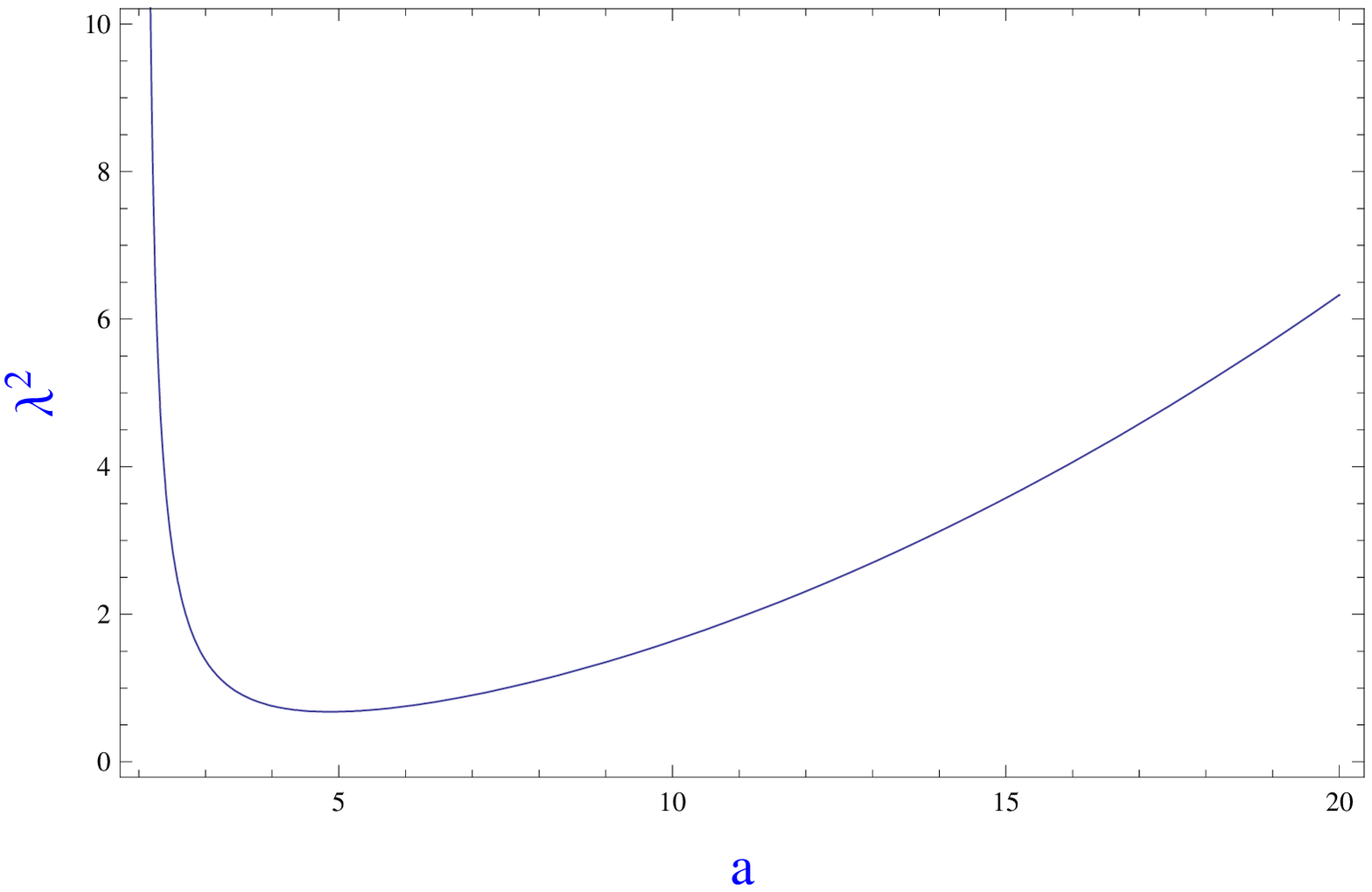}
\caption{\label{Fig4} The eigenvalue $\lambda^2$ against   $a$ with
Plank unit  and $V=0.012$. The left and right panels show results of
$a<a_*$ and $a>a_*$, respectively. }
\end{figure}

If one obtains the equilibrium points, their stabilities are
determined by the eigenvalues coming from linearizing the system
near these points. By calculation, we get the eigenvalue $
\lambda^2$, but the results are very tedious. We will not give them
here. If $\lambda^2<0$ the corresponding equilibrium point is a
center otherwise it is a saddle.  Since $\lambda^2$ is positive for
$a>a_*$ and negative for $a<a_*$, as shown in Fig.~(\ref{Fig4}),
equilibrium points occurring in the classical region are saddles and
those in the semiclassical region are centers. Therefore from
Fig.~(\ref{Fig2}) we obtain that in LQC there is a center
equilibrium point in the semiclassical region, which originates from
the loop quantum effects, and a saddle point in the classical
region, which is similar to the classical  general relativity.

\begin{figure}[htbp]
\includegraphics[width=7cm]{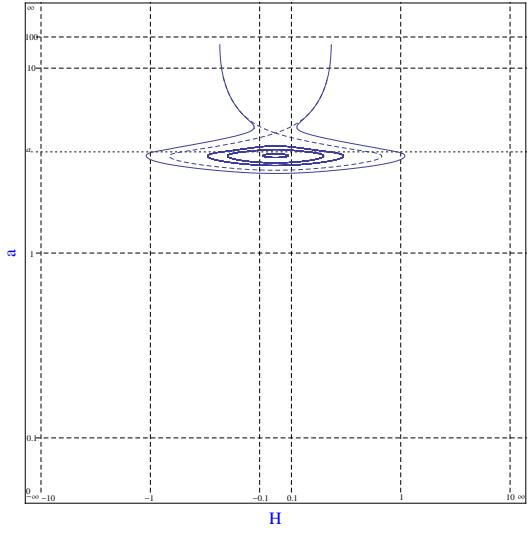}
 \caption{\label{Fig5} This figure  represents the phase diagram corresponding to
Fig.~(\ref{Fig2}) with the Plank unit, $J=100$ and $\alpha=1$. The
axes have been compactified  using the relations $x(t)=\arctan(H)$
and $y(t)=\arctan(\ln a)$}
\end{figure}

In Fig.~(\ref{Fig5}) we give the phase portraits for demonstrating
the center equilibrium point and the saddle point. The separatrix is
represented by the dashed line. Apparently if the value of the
potential is less than $V_{crit}$ the universe can stay at a stable
state past-eternally and then undergo an infinite oscillation with
the tachyon field climbing up its potential. Once the potential
reaches the critical value $V_{crit}$, the oscillating phase is
broken and then the universe enters into an inflationary era.

\section{Conclusion}
In summary, in LQC with the inverse volume corrections to the
tachyon field and the gravity, there are two equilibrium points. The
one appearing in the classical region is a saddle point as that in
general relativity; the other in semiclassical region is a center
which comes from the loop quantum effects. If the potential of
tachyon field asymptotes to a positive constant and is less than
$V_{crit}$ as $t\rightarrow -\infty$, the universe can stay
past-eternally at a stable state. With the tachyon field climbing up
its potential slowly, the universe will undergo non-singular
oscillations around the center equilibrium point. Once the potential
reaches the critical value $V_{crit}$, the saddle point coincides
with the center point and then the cycles are broken by the
emergence of an inflation. Thus the universe enters into a de Sitter
expansion phase. In addition we find that,  in order to obtain a
successful emergent tachyon inflation, the allowed region of
$\dot{\phi}^2$ at the stable state is constrained. For the normal
scalar field the emergent inflation will be studied in the future.

\begin{acknowledgments}
We appreciate very much the insightful comments and help suggestions
by anonymous referee. P. Wu is partially supported by the National
Natural Science Foundation of China  under Grant No. 10705055, the
Scientific Research Fund of Hunan Provincial Education Department,
the Hunan Provincial Natural Science Foundation of China under Grant
No. 08JJ4001, and the China Postdoctoral Science Foundation. S. N.
Zhang acknowledges partial funding support by the Yangtze Endowment
from the Ministry of Education at Tsinghua University, Directional
Research Project of the Chinese Academy of Sciences under project
No. KJCX2-YW-T03 and by the National Natural Science Foundation of
China under grant Nos. 10521001, 10733010, 10725313, and by 973
Program of China under grant 2009CB824800. H. Yu is supported in
part by the National Natural Science Foundation of China under
Grants No.10775050 and the SRFDP under Grant No. 20070542002,
\end{acknowledgments}

\end{document}